\documentclass[namedreferences]{SolarPhysics}
\usepackage[optionalrh]{spr-sola-addons} 
\usepackage{graphicx}                    
\usepackage{color}                       
\usepackage{url}                         
\usepackage{lscape}       
\usepackage{rotating}

\usepackage{solaheader} 
\newcommand{\solareceiveddate}{28 August 2012} 
\newcommand{\solaaccepteddate}{22 December 2012}
\newcommand{\solaonlinedate}{23 December 2012}
\newcommand{\Xdoi}{s11207-012-0220-5}  
\begin{document}              
\begin{article}              
\begin{opening}              
\title{Cyclic and Long-term Variation of Sunspot Magnetic Fields}
              
\author{Alexei A. \surname{Pevtsov}$^{1}$\sep
Luca \surname{Bertello}$^{2}$\sep
Andrey G. \surname{Tlatov}$^{3}$\sep
Ali  \surname{Kilcik}$^{4}$\sep
Yury A. \surname{Nagovitsyn}$^{5}$\sep
Edward  W. \surname{Cliver}$^{6}$\sep}                            
\runningauthor{Pevtsov et al.} 
\runningtitle{Cyclic and Long-term Variation of Sunspot Magnetic Fields}
                            
\institute{$^1$ National Solar Observatory, Sunspot, NM 88349, 
U.S.A.  email: \url{apevtsov@nso.edu}\\
$^2$ National Solar Observatory, 950 N. Cherry Avenue, 
Tucson, AZ 85719, U.S.A.
email: \url{lbertello@nso.edu}\\ 
$^3$ Kislovodsk Solar Station of Pulkovo Observatory, PO Box 145,
Gagarina Str., 100, Kislovodsk, 357700 Russian Federation 
email:\url{tlatov@mail.ru}\\
$^4$ Big Bear Solar Observatory 40386 Big Bear City, CA, U.S.A.\\ 
email: \url{kilcik@bbso.njit.edu}\\ 
$^5$ Pulkovo Astronomical Observatory, Russian Academy of Sciences,
Pulkovskoe sh. 65, St. Petersburg, 196140 Russian Federation 
email: \url{nag@gao.spb.ru}\\
$^6$ Space Vehicles Directorate, Air Force Research Laboratory, Sunspot, 
NM 88349, U.S.A. 
email: \url{ecliver@nso.edu}} 

\begin{abstract}       
Measurements from the Mount Wilson Observatory (MWO) are used to study
the long-term variations of sunspot field strengths from 1920 to 1958.
Following a modified approach similar
to that in \inlinecite{Pevtsov2011}, for  each observing week we
select a single sunspot with the strongest field strength measured
that week and then compute monthly averages of these weekly maximum
field strengths. The data show the solar cycle variation of the peak
field strengths with an amplitude of about 500-700 gauss (G), but no
statistically significant long-term trends.  Next,  we use the sunspot
observations from the Royal Greenwich Observatory (RGO) to establish
a relationship between the sunspot areas and the sunspot field
strengths for Cycles 15-19. This relationship is then used to create a
proxy of peak magnetic field strength based on sunspot areas
from the RGO and the  USAF/NOAA network for the period from
1874 to early 2012. Over this interval, the magnetic field proxy
shows a clear solar cycle variation with an amplitude of 
500-700 G and a weaker long-term trend. From 1874 to around 1920,
the mean value of magnetic field proxy increases by about 300-350
G, and, following a broad maximum in 1920-1960, it
decreases by about 300 G. Using the proxy for the magnetic field
strength as the reference, we scale the MWO field measurements to the
measurements of the magnetic fields in \inlinecite{Pevtsov2011} to
construct a combined data set  of maximum  sunspot field
strengths extending from 1920 to early 2012. This combined data set
shows strong solar cycle variations and no  significant long-term
trend (linear fit to the data yields a slope  of $-0.2\pm$0.8 G year$^{-1}$). On
the other hand, the peak sunspot field strengths  observed at the
minimum of the solar cycle show a gradual decline over  the last
three minima (corresponding to cycles 21-23) with a mean downward
trend of  $\approx$ 15 G year$^{-1}$.

\end{abstract}                                                 
                                                                      
\end{opening}

\section{Introduction}

Recently, several studies have concentrated on the long-term
variations of field strengths in sunspots. The question at the core of
these investigations is: "Has the strength of sunspot magnetic fields
gradually declined over the last two cycles?"

Penn and Livingston (2006, 2011) measured the field strength using the
separation of two Zeeman components of the magnetically sensitive
spectral line Fe {\sc i} 1564.8 nm over the declining phase of solar cycle
23 and the rising phase of cycle 24. The measurements were taken on a
daily basis (in quarterly observing intervals due to telescope
scheduling). In early years, only the large sunspots were measured;
more recent observations are aimed at including all sunspots and pores
that are present on the disk. Monthly averages of these measurements
show a gradual decrease in sunspot field strength from the beginning
of the project (late 1998) to the present (mid-2012).  It is possible
that the non-uniformity of the data set ({\it i.e.}, fewer measurements at
the beginning, newer observations include both sunspots and pores) may
result in such a decline in average field strengths.  However, the
most recent study by \inlinecite{Livingston2012} indicates that there
is no change in the shape of the distribution of measured field
strengths over the observing period; only the mean of the distribution
changes.  This constancy in the shape of the field distribution
appears to rule out the  speculation that the decline in field
strengths could be explained by the non-uniformity of the data set.

\inlinecite{Watson2011} employed the magnetic flux measurements from
the Michelson Doppler Imager (MDI) on board the {\it Solar and Heliospheric
Observatory} (SOHO), and studied the magnetic flux changes in sunspots
over cycle 23 (1996-2010). As the MDI measures only the
line-of-sight fluxes, they reconstructed the vertical flux under the
assumption that the magnetic field in sunspots is vertical.  The
results showed a solar cycle-like variation and only a minor long-term
decrease in vertical  magnetic flux of the active regions.
\inlinecite{Pevtsov2011} employed the manual measurements of magnetic
field strengths  taken in Fe {\sc i} 630.15-630.25 nm  wavelengths in the period of  
1957--2010 in seven observing stations forming the solar observatory
network across eleven time zones in what is now  Azerbaijan, Russian
Federation, and Ukraine. To mitigate the differences in the
atmospheric seeing and the level of the observers'  experience,
\inlinecite{Pevtsov2011} considered only the daily strongest sunspot
measurements  for each day of observations. The sunspot field
strengths were found to vary strongly with the solar cycle, and no
long-term decline was noted. \inlinecite{Rezaei2012} used the
magnetographic observations from the Tenerife Infrared Polarimeter  at
the German Vacuum telescope in the period of 1999-2011 to confirm the cycle
variations of sunspot magnetic fields. Comparing maximum  field
strengths measured at the rising phases of cycles 23 and 24, the
authors had noted a slight reduction in field strengths at the
beginning of cycle 24. Still, the main variations in the magnetic
field strength  were found to be related to the solar cycle.

With the exception of \inlinecite{Pevtsov2011}, all previous studies
based their conclusions on the data from the most recent full cycle
23. In our present article, we extend the analysis to earlier solar
cycles. First, we use the data from the Mount Wilson Observatory (MWO)
and apply the same technique as in \inlinecite{Pevtsov2011}. Next, we
establish a correlation between the area of a sunspot and its field 
strength, and we use this correlation to construct a proxy for the
magnetic field strength based on areas of sunspots measured by the Royal
Greenwich Observatory (RGO) and USAF/NOAA. The proxy for the  magnetic
field strength  allows us to extend the analysis to cycles 11-24 and
to scale the 1920-1958 MWO peak field measurements to those of
\inlinecite{Pevtsov2011} for the 1957-2010 interval.  Our analysis is
presented in Sections 2 and 3, and our results are  summarized and
discussed in Section 4.

\section{Sunspot Field Strength Measurements from Mount Wilson Observatory 
(1920-1958)}
       
To investigate the properties of the sunspot magnetic fields, we
employ the observations from the MWO from
May 1920 till December 1958. Specifically, we select only the data
published in the {\it Publications of the Astronomical Society of the
Pacific} (PASP).  Although the sunspot field strength measurements
continued to the present  days (with a major interruption between
September 2004 to January 2007 due to funding problems), their regular
publication was discontinued at the end of 1958.  This 1920-1958 part
of the MWO data set should be considered as the most uniform; in later
years there were several major changes to the instrumentation and the
observing procedure ({\it e.g.}, multiple replacements of the spectrograph
grating, selection of a new spectral line for measurements, and a
different tilt-plate).  In addition, the resolution of the field
strength measurements changed from 100 gauss (G) through 1958 to 500 G
for the latter measurements.  A summary of these changes can be found
in \inlinecite{Livingston2006}.

We digitized the data summaries in PASP and verified the newly
constructed tables against the published record.  In a few instances,
the MWO had issued corrections to the original tables (also published
in PASP). All these corrections were taken into account in the process
of data verification.  Finally, we found a small number of
inconsistencies in the original tables, which were also corrected.

The MWO sunspot summaries provide MWO sunspot number, its field \\
strength (no polarity information), latitude, and an estimated date of
the central meridian passage (later data also show the first and last
days of an active region on the disk as well as its magnetic
classification). In this study, we only use the sunspot field
strength, latitude, and the date of the central meridian crossing.  For
some active regions the maximum field strength is estimated (see
explanation in PASP, {\bf 50}, 61-64, 1938:
``When it seems probable that the greatest field-strength observed in
any group was not the maximum for that group, an estimated value is
given in parentheses").  The measurements of field strength were made
manually using a glass  tilt-plate.  By tilting the plate, the
observer would co-align two Zeeman components of the spectral line, and
the tilt angle will be translated to a linear  (wavelength)
displacement. Since the stronger field strengths require larger  tilt
angles, this procedure may introduce a non-linearity in the relation
between the plate's tilt and the linear displacement. This slight
non-linearity was corrected following the procedure specified in
\inlinecite{Livingston2006}. Only  measurements above 2400 G were
affected by this non-linearity.

A quick examination of the entire data set revealed several systematic
effects. First, we found an increased number of values in parentheses
(estimated values) in later years of the data set as compared with
earlier years.  Thus, we excluded all estimated data from our
investigation.  Second, we noted a gradual increase in the fraction of
measurements with weak field strengths from earlier to later years.
For example, for observations taken during cycle 15 only 4\% of all
measurements have field strengths of 100 G. For later cycles, the
fraction increases significantly to 9\% (cycle 16), 15\% (cycle 17),
14\% (cycle 18), and 18\% (cycle 19).  The tendency for a larger
fraction of  measurements with weaker fields is quite obvious in the
annual number of measurements with zero fields (when observations were
taken, but the fields were considered to be weaker  than 100
G). Even after normalizing for the level of sunspot activity
(using international sunspot numbers), the annual number of
measurements with zero fields shows a steady (and significant)
increase from 1920 to 1958.
 
Figure 1 shows the fractional distribution of sunspot field strengths
in 400 G bins in the range of 0-5000 G. Data for different
cycles are shown by different colors. The most significant differences
are confined to first two bins. For field strengths in the range of
100--400 G, there is a systematic increase in the fraction of weak
field measurements from about 8\% in cycle 15 to about 25\% in cycle
19.  The above  fractions are given in all the measurements
taken in a given cycle.  The measurements in the next bin (500--800
G) show the opposite trend with a  decrease in the fraction of
measurements from about 22\% in cycles 15 and 16 to about 14\% in
cycle 19. Stronger fields do not show any systematic behavior from one
cycle to the next.

One can speculate that at least some of these systematic effects could
be due to changes in the observing requirements ({\it e.g.}, increase in
``granularity" of measurements, when measurements are taken in separate
umbrae of a single sunspot) and/or a ``learning curve" effect (when
with increased experience, the observer starts expanding the
measurements to smaller sunspots). An increase in the scattered light
might have similar effects on the visual measurements of the magnetic
fields.  These systematic effects may affect the average value of the
field strengths. For example, including the larger number of
measurements with the smaller field strengths will decrease the
average value. To mitigate the negative effects of this possible
change in statistical properties of the data sample, we employ an
approach  similar to the one used by
\inlinecite{Pevtsov2011}, where only the strongest measured  field
strength is selected for any given day of observations.

Next, we investigate the temporal behavior of the strongest sunspot
field strengths.  Since the published summaries of MWO observations do
not provide the date of measurements, we adopted the estimated date of
the central meridian crossing as a proxy for the day of observations.

The application of \inlinecite{Pevtsov2011} approach to
MWO data is complicated by the fact that MWO measurements are, 
strictly speaking, not the daily observations. In many cases, there are
daily drawings,  but no corresponding magnetic field
measurements. Also, when measurements do exist, they may exclude some
(sometimes even the largest) sunspots that were present on the
disk. As one example, during a disk passage of a large sunspot (MWO AR
7688) on 17 -- 28 November 1944, the measurements of magnetic field
were taken only on 17 -- 19 November and 22 November. On 22 January 
1957, the field strengths were measured only in  smallest sunspots and
pores; the largest sunspots were not measured. From 27 February -- 2 March
1942, the measurements alternate between the largest sunspots on the
disk (one day) and the smallest sunspots (the following day).  Since the
published summaries of MWO observations provide only the  estimated
date of the central meridian crossing (but not the date of
observations), the selection of only one measurement of the strongest
field strength for any given day of the observations as in
\inlinecite{Pevtsov2011} may not work well ({\it i.e.}, measurements of
weaker field strengths are less likely to be excluded even if there
are sunspots with stronger fields on the disk).  Thus, we modified the
method by selecting only the measurement of the  strongest field
strength for any given week of observations.

\inlinecite{Livingston2006} suggested that, since for the field
strengths less than 1000 G, the Zeeman splitting for Fe 617.33 nm
becomes comparable to the Doppler width of the spectral line, the
measurements of such fields are unreliable.  That said, a well-trained
observer can consistently measure fields weaker than 1000 G. The
reader is referred to \inlinecite{Livingston2006} for one example when
multiple measurements from the same observer agree within 10 G. We
also examined several drawings in more detail and found a good
persistence in day-to-day measurements of pores with field strengths
below 1000 G. Nevertheless, to be cautious, we chose to exclude field
strength measurements below 1000 G in our analysis.  We note, however,
that inclusion of field strength measurements below 1000 G does not
change the main results of this article.

Figure 2 (upper panel) shows the latitudinal distribution of active
regions in our MWO data set and the monthly averages of the (weekly)
strongest field strengths (Figure 2, lower panel).  The most prominent
trend in the data is a solar cycle variation, with minimum field
strength in sunspots around the minima of solar cycles, and maximum
field strength near the maxima of solar cycles (see, Table \ref{t1}).
To verify the presence (or absence) of the long-term trend, we fitted
the data with linear and quadratic functions. A linear fit suggests a
negligible decrease in sunspot field strengths over 40 years of about
0.8$\pm$1.7 G year$^{-1}$.

\section{Sunspot Area as Proxy for Magnetic Field Strength}
       
Several studies (e.g., \opencite{Houtgast1948}; \opencite{Rezaei2012})
found a correlation between the area of sunspots ($S$, in millionths of
solar hemisphere, MSH) and their recorded field strength ($H_{\rm MAX}$).
\inlinecite{Ringnes1960} found the strongest correlation between the
logarithm of area and the field strength. Here, we use this
relationship to investigate the changes in sunspot magnetic field
strength over a long time interval by employing the area of sunspots as
a proxy for magnetic field strength. Figure 3 shows the relationship 
between the logarithm of the area of sunspots  and their field
strength. Areas are taken from the independent data set of the 
RGO.  For this plot, we established a
correspondence between the active regions in the MWO and RGO data
using the date of central meridian passage and the latitude of active
regions. Only the regions that had a small difference ($\le 0.6^\circ$)
in latitude and less than 4.8 h (0.2 day) in the time of
the central meridian crossing between the two data sets were
included. We also excluded RGO areas smaller than 10
MSH.  Solar features with area $S \le$ 10 MSH are
small spots or pores and their field strengths are more likely to have
large errors in the measurements.

Similar to the previous studies, we find a reasonably good correlation
between the (logarithm of) sunspot areas and the magnetic
field strength (Pearson's correlation coefficient $\rho$=0.756). Similar
(strong) correlation coefficients were found for individual cycles
(see Table 2). To verify the statistical significance of the
correlations, we used the $t$-test; the $t$-values and the cutoff value for
99\% confidence level are shown in Table 2. Since the $t$-values are
well above the 99\% cutoff  values, all the found correlations are
statistically significant.  The relationship between the
magnetic field strength and the logarithm of area of sunspots  is very
similar for cycles 15-17. For cycle 18 the relation is ``steeper"
(sunspots with the same areas correspond to stronger magnetic fields
as compared with  cycles 15-17), and for cycle 19, the field-area
relationship is  somewhat weaker and  more similar to the relationship found 
in cycle 15.   Our data set only partially includes these two
cycles. Cycles 16-18 are included in their entirety.  Overall, we find
that the MWO data for cycle 18  contain a slightly larger fraction of
stronger field measurements, as  compared with all other cycles ({\it e.g.},
see Figure 1).  Limiting the data to three complete cycles 16-18 does
not significantly affect the coefficients for the functional relation
between sunspot area  and magnetic field strength (see Table 2).
Figure 3 indicates a non-linearity between the   area of
sunspots and their magnetic field strength. On the other hand, a
quadratic fit to the data (solid line, Figure 3) suggests  that this
non-linearity is small. For simplicity, in the following discussion we
employ a linear relation between the logarithm of sunspot
areas  and their magnetic field strength.

Using the relation $H_{\rm MAX} = -774.2 + 536.0 \times \ln(S)$ based on
cycles 15-19 (Table 2), we created a proxy for the magnetic field
strength based on the sunspot areas.  The data used for this exercise
represent a combination of the RGO measurements from   1874--1977 and
the USAF/NOAA data from 1977--early 2012. The USAF/NOAA data  were
corrected (by D. Hathaway) by a factor of 1.4 as described in
http://solarscience.msfc.nasa.gov/greenwch.shtml.   Figure 4 shows
annual averages of the proxy of the magnetic field derived from the
deprojected active region areas. For consistency with the scaling
between the logarithm of area and the magnetic field strength derived earlier,
Figure 4 does not include areas smaller than 10 MSH. 
We note, however, that inclusion of all areas does not
noticeably change Figure 4.  Similar to the direct measurements of the
magnetic field, the proxy also exhibits the solar cycle variations
with an approximate amplitude of 500-700 G. The data also show a
long-term trend (see a parabolic fit to the data), with the mean value
of the field proxy increasing from 1874 to around 1920 by about 300 G
and then, following a broad relatively flat maximum to around 1960,
decreasing by about same amount by early 2012. For comparison, Figure
4 also shows the magnetic field proxy computed using a
quadratic  fit to the area-field strength relation (red line).

Finally, using the magnetic field proxy as the reference, we combined
the MWO data and the Russian sunspot field strength observations from
\inlinecite{Pevtsov2011} into a single data set. Figure 5 shows two
data sets on the same scale. MWO data were re-scaled to Russian data
using the functional dependencies between the $H_{\rm MAX}$ (MWO) and
magnetic field proxy (sunspot areas) and similar dependency for the
Russian data.  With the scaling,  the mean level of approximately 2500
G in Figure 5  is in agreement with  that observed at NSO/Kitt Peak
from 1998-2005 \inlinecite{Penn2006}.

The most prominent feature of the combined data shown in Figure 5 is
the cycle variation of sunspot field strength. Maxima and  minima in
the sunspot field strength are in relative agreement with maxima and
minima in the sunspot number.  The combined data set (Figure 5) shows
no statistically significant long-term trend. A linear fit to the data
(not shown in Figure 5) has a slope of $\approx -1.9\pm 0.8$
G year$^{-1}$.  This trend appears to be entirely due to the points
corresponding to the deep minimum around 2010, and it disappears once
these few points are excluded from the fitting (the fit shown in
Figure 5 has a slope of $\approx -0.2\pm 0.8$ G year$^{-1}$). In either
case, the amplitude of the trend is significantly smaller than the
$-52$ G year$^{-1}$ reported by \inlinecite{Penn2006}. As the
\inlinecite{Penn2011} data cover the period of 1998 -- 2011, the trend
in the field strengths found in their data may be dominated by the
declining phase of solar cycle 23.

\section{Discussion}

In this study we employed the observations of the sunspot field strengths 
from a subset of the MWO data covering solar cycles 15-19 (1920--1958). The 
data were analyzed using a modified approach of \inlinecite{Pevtsov2011}, 
where a single strongest field measurement was selected for each week of 
observations. This approach allows one to mitigate the 
effects of some negative properties of the data set, for example, a
systematic increase in the number  of weak field measurements from the
beginning of the data set to its end.  Our findings confirm the
presence of the 11-year cycle variation in the  sunspot  daily
strongest field strengths similar to that found in the previous studies
for solar Cycles 19-23 (\opencite{Pevtsov2011}, \opencite{Watson2011},
\opencite{Rezaei2012}). On the other hand, no significant secular
trend is found for the period covered by the MWO data (cycles 15-19).
We do see a weak gradual decrease in average field strengths of  
$\approx$2.1$\pm$0.9 G year$^{-1}$.

A correlation between the area of sunspots and the magnetic field
strength allows us to use the sunspot area as a proxy for the magnetic
field strength. Using this approach, we show that the cycle variations
(in the magnetic field strengths as represented by their proxy) are
present during cycles 11-24 (1874--early 2012). The amplitude of these
cycle variations is about 1000 G between the solar activity minima and
maxima. The magnetic field strength proxy does show a secular
trend: between 1874 and $\approx$ 1920, the mean value of the magnetic
field proxy increased by about 300 G, and following a broad
maximum in 1920-1960, it decreased by 300 G. The nature of
this trend is unknown, but we note that the broad ``maximum" in
1920-1960 includes the mid-20th century maximum of the current
Gleissberg cycle, which began at about 1900.  The long-term trend in
magnetic field strength proxy based on sunspot area (Figure 4) could
be subject to several uncertainties. For example, sunspot area
estimates from the RGO observations prior to $\approx$ 1910 appear to be
systematically low, in  comparison with later data (Leif Svalgaard,
private communication).  Also, the functional dependency between
the sunspot area and the peak magnetic field strength may be
non-linear and/or  vary from one cycle to another (see
the quadratic and linear fits  in Figure 3). For example, a long-term
variation in the  $H_{\rm MAX}- S$ dependency was found in
\inlinecite{Ringnes1960}. The steepness of linear
regression  in cycle $n$ may correlate with the amplitude of cycle
($n+1$) (compare  the $B$ coefficients and the yearly international sunspot
number, SSN$_{n+1}$, in  Table 2) although our statistical sample  is
too small to be more conclusive with respect to this possible
dependency.  Nevertheless, the $H_{\rm MAX}$ proxies computed using
linear and non-linear scaling functions agree reasonably
well (compare black and red lines in Figure 4).

Using the proxy of $H_{\rm MAX}$ as a reference allows us to combine the
MWO and Russian measurements of the magnetic field into a single data
set. Although this data set does not show any statistically
significant long-term trend comparable in amplitude with the trend
reported in \inlinecite{Penn2011}, one can notice that the last three
(see red curve in Figure 5) minima in the sunspot field strengths get
progressively lower. The overall trend derived from this tendency is
about 15 G year$^{-1}$, which could be associated with the
Gleissberg  cycle variations as derived from other
studies  ({\it e.g.}, \opencite{Mordvinov1999}).  Alternatively, this could
be an indication of some other long-term pattern in sunspot field
strengths. One can note a similar tendency for progressively lower
field strengths in the solar minima of cycles 12-14 (Figure 4) and cycles
17-19 (Figures 4 and 5). We leave the investigation of these patterns
to future studies.

\begin{acks} Y. N. and  A. T. acknowledge a partial support from the Federal
Program "Scientific and Pedagogical Staff of Innovative Russia",  the
Russian Foundation for Basic Research (grants No. 10-02-00391 and
12-02-00614), and the Programs of the Presidium of Russian Academy of
Sciences No. 21 and 22.  AAP acknowledges funding from NASA's
NNH09AL04I inter agency transfer.  National Solar Observatory (NSO) is
operated by the Association of Universities for Research in Astronomy,
AURA Inc under cooperative agreement with the National Science
Foundation (NSF).

\end{acks}

\clearpage 

\begin{figure} 
\includegraphics[width=1.0\textwidth,clip=]{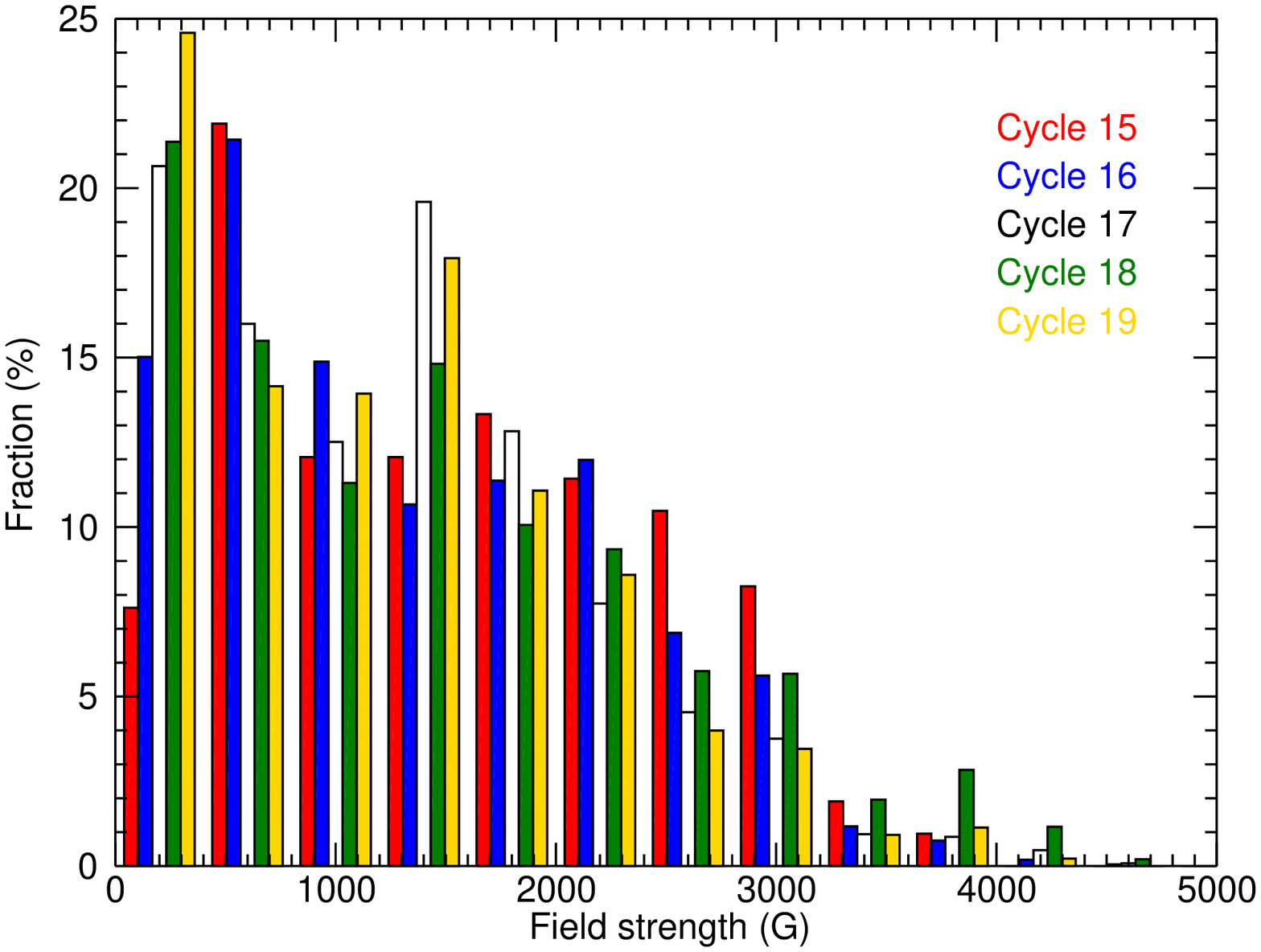}
\caption{Fractional distribution of sunspot field strengths for
different cycles. The bin size is 400 G. Measurements with zero field
strength are excluded.  
\label{f2}} 
\end{figure} 

\begin{figure}
\includegraphics[width=1.0\textwidth,clip=]{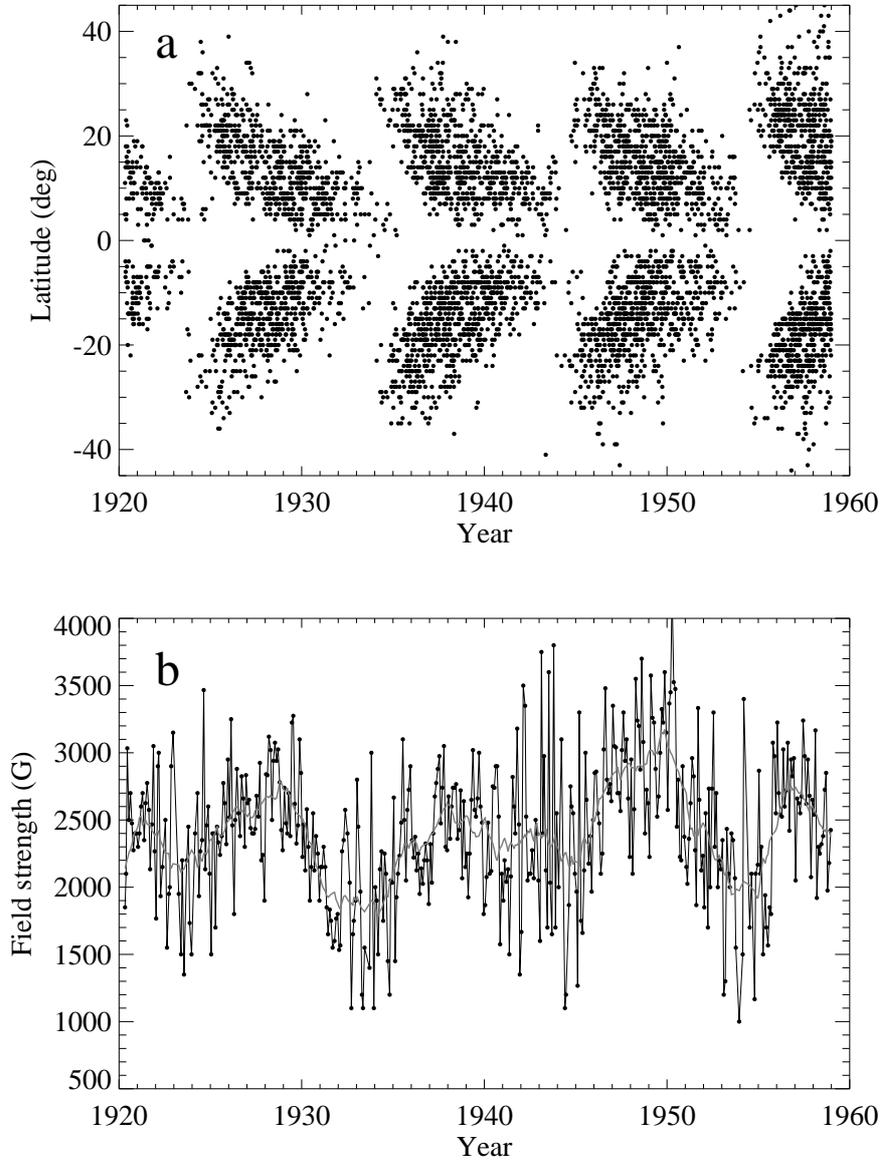}
\caption{Time-latitude distribution of sunspots in MWO data set
included in our study (upper panel), and monthly average of daily peak sunspot
field strengths (lower panel, dots connected by thin line). The thick grey
line is 18-point running average.  
\label{f1}} 
\end{figure} 
\begin{figure}
\includegraphics[width=1.0\textwidth,clip=]{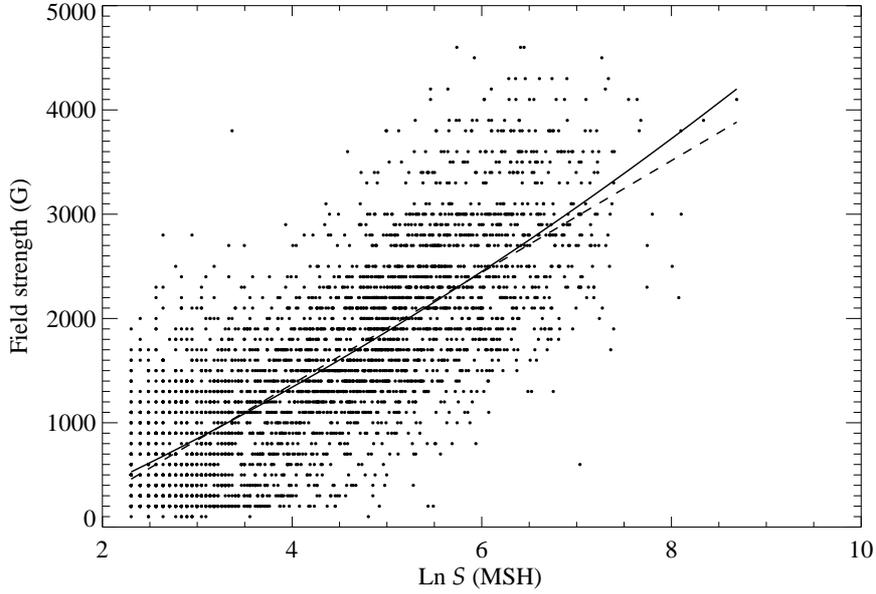} 
\caption{Magnetic field strength (from MWO observations) vs. the 
natural logarithm of sunspot
area (from RGO observations) for cycles 15-19. The dashed line is a
first degree polynomial fit to the data. Fitted coefficients are shown
in Table \ref{t2}, in entry for ``all cycles". 
For comparison, the solid
line shows a fit by a quadratic function.
\label{f3}}
\end{figure} 

\begin{figure} 
\includegraphics[width=1.0\textwidth,clip=]{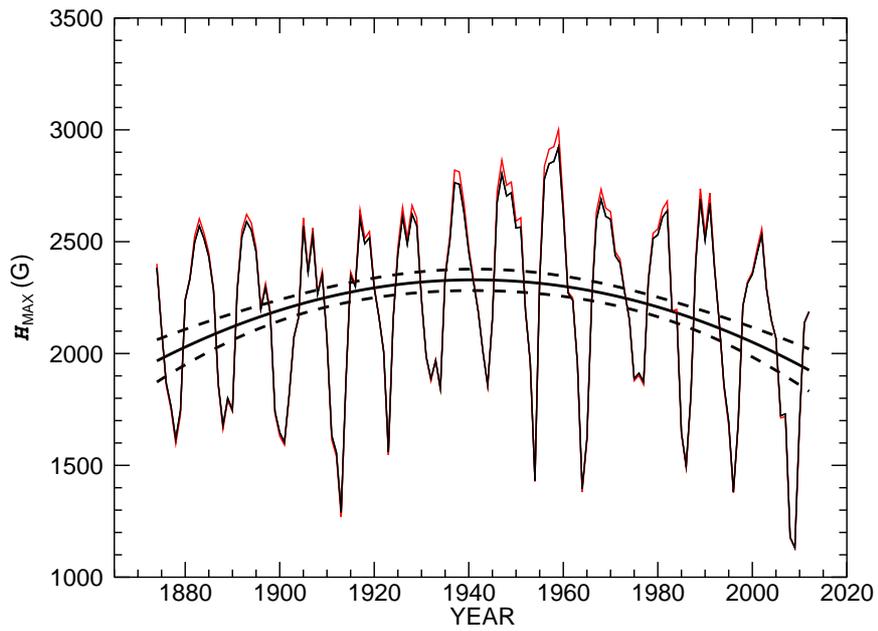}
\caption{Proxy of the magnetic field strength computed from the deprojected
sunspot areas. Annual averages are shown as a thin solid line. The thick
solid line is a second degree polynomial approximation to the data, and
two thick dashed lines represent a one-standard deviation error band for
the fit. The red line shows the magnetic field proxy derived using the 
quadratic dependency shown in Figure 3.   
\label{f4}} 
\end{figure} 

\begin{figure} 
\includegraphics[width=1.0\textwidth,clip=]{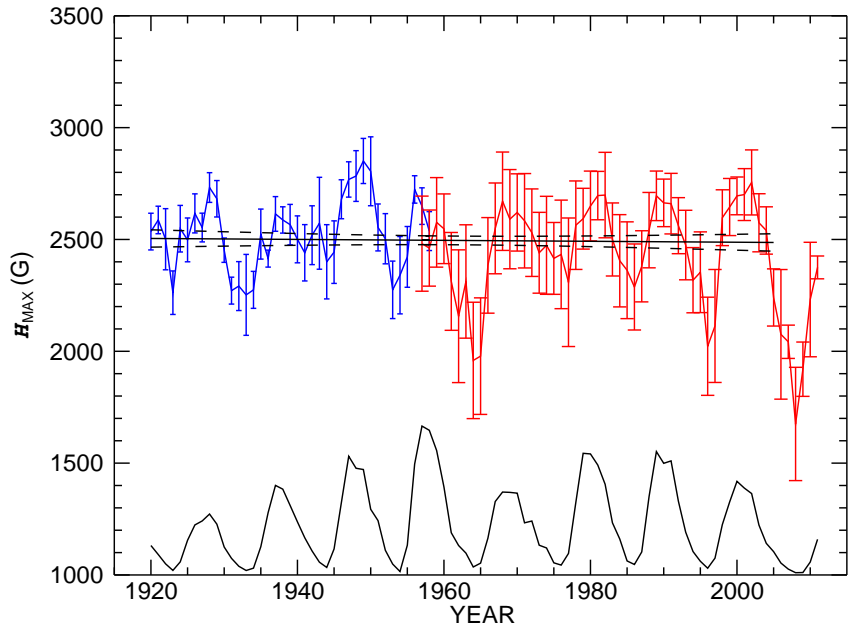}
\caption{Annual averages of the magnetic field measurements from the
MWO  (blue color, up to 1958) and Russian data set from Pevtsov {\it et al.} 
(2011, red).  MWO data are scaled to Russian data set using proxy of magnetic
field  strength  in Figure 4 as the reference. Error bars show a
$\pm\ 1 \sigma$ standard  deviation of mean values. The thick black
straight line is a linear approximation  to the MWO and the Russian
data. Dashed  lines indicate the statistical uncertainties of this
linear approximation.  The black line in the lower part of the figure
shows the annual international  sunspot numbers.
\label{f5}}
\end{figure}

\begin{table} 
\caption{Years of maxima and minima for solar cycle and sunspot field 
strengths}
\label{t1} 
\begin{tabular}{cccc} 
\hline \multicolumn{2}{c}{Solar
cycle}&\multicolumn{2}{c}{Sunspot Field}\\
Minimum&Maximum&Mminimum&Maximum\\ 
\hline 
1923& &1923.6&\\ &1928& &1928.8\\
1933& &1933.3&\\ & 1937& &1938.1\\ 
1944& &1944.7&\\ 
&1947& &1949.8\\ 
1954& &1953.9&\\ 
\end{tabular} 
\end{table} 

\begin{table}
\caption{Correlation ($\rho$) and fitted coefficients for $H = A + B \times
\ln(S)$ dependency between magnetic field strength ($H$) and deprojected 
area ($S$) of an active region. Student's $t$-values and maximum 
sunspot  number (SSN) for the $(n+1)$-th cycle are also shown.}
\label{t2} 
\begin{tabular}{ccccccc} 
\hline Cycle
No.&$A$&$B$&$\rho$&$t$-value&99\%-level&SSN$_{n+1}$\\ \hline 
Cycle 15&-274.1$\pm$177.6&507.3$\pm$40.2&0.775&12.633&2.623&78.1\\
Cycle 16&-475.1$\pm$63.4&514.9$\pm$13.9&0.811&36.947&2.583&119.2\\ 
Cycle 17&-771.0$\pm$59.9&523.2$\pm$13.2&0.781&39.595&2.581&151.8\\ 
Cycle 18&-1106.9$\pm$78.9&609.2$\pm$16.9&0.739&35.966&2.580&201.3\\ 
Cycle 19&-800.4$\pm$69.5&495.4$\pm$14.9&0.784&33.252&2.583&110.6\\ 
All cycles&-774.2$\pm$35.6&536.0$\pm$7.7&0.756&69.170&2.577\\ 
Cycles 16-18&-806.3$\pm$41.0&551.7$\pm$8.9&0.761&61.670&2.578\\ 
\end{tabular} 

\end{table}

\end{article} 

\end{document}